\documentclass[10pt, conference]{IEEEtran}

\IEEEoverridecommandlockouts

\usepackage{amsmath,amssymb,amsfonts}
\usepackage{graphicx}
\usepackage{textcomp}

\usepackage{optidef}
\usepackage[nocompress]{cite}

\usepackage{rotating}
\usepackage{multirow}
\usepackage{array}
\usepackage{xcolor}
\usepackage{url}
\usepackage{pgfplots}
\usepackage{verbatim}
\usepackage{tikz}
\usepackage{algorithm}
\usepackage{algorithmic}
\usepackage{breqn}
\usepackage{amsmath}
\usepackage{aurical}
\usepackage[T1]{fontenc}

\usepackage[utf8]{inputenc}
\usepackage{nomencl}
\usepackage[normalem]{ulem}
\usepackage{soul}
\makenomenclature

\usepackage{subfig}

\newcommand{\tabmore}{\hspace{10pt}}

\newlength\myindent
\newcommand\bindent{%
  \begingroup
  \setlength{\itemindent}{\myindent}
  \addtolength{\algorithmicindent}{\myindent}
}
\newcommand\eindent{\endgroup}

\usepackage{mathtools}

\usepackage{color, colortbl}
\definecolor{LightCyan}{rgb}{0.88,1,1}

\def\BibTeX{{\rm B\kern-.05em{\sc i\kern-.025em b}\kern-.08em T\kern-.1667em\lower.7ex\hbox{E}\kern-.125emX}}

\usetikzlibrary{mindmap,shadows,backgrounds}
\graphicspath{{figures/}}
\pgfplotsset{compat=1.14}

\def\BibTeX{{\rm B\kern-.05em{\sc i\kern-.025em b}\kern-.08em
    T\kern-.1667em\lower.7ex\hbox{E}\kern-.125emX}}

\ifCLASSINFOpdf

\else

\fi

\begin{document}

\title{Machine Learning-based Inter-Beam Inter-Cell Interference Mitigation in mmWave}

\author{\IEEEauthorblockN{Medhat Elsayed, Kevin Shimotakahara and Melike Erol-Kantarci, \IEEEmembership{Senior Member, IEEE}} \\
\IEEEauthorblockA{School of Electrical Engineering and Computer Science \\ University of Ottawa \\ Ottawa, Canada, \\
Email: \{melsa034, kshim017, melike.erolkantarci\}@uottawa.ca}
}

\maketitle

\begin{abstract}
In this paper, we address inter-beam inter-cell interference mitigation in 5G networks that employ millimeter-wave (mmWave), beamforming and non-orthogonal multiple access (NOMA) techniques. Those techniques play a key role in improving network capacity and spectral efficiency by multiplexing users on both spatial and power domains. In addition, the coverage area of multiple beams from different cells can intersect, allowing more flexibility in user-cell association. However, the intersection of coverage areas also implies increased inter-beam inter-cell interference, i.e. interference among beams formed by nearby cells. Therefore, joint user-cell association and inter-beam power allocation stand as a promising solution to mitigate inter-beam, inter-cell interference. In this paper, we consider a 5G mmWave network and propose a reinforcement learning algorithm to perform joint user-cell association and inter-beam power allocation to maximize the sum rate of the network. The proposed algorithm is compared to a uniform power allocation that equally divides power among beams per cell. Simulation results present a performance enhancement of $13-30\%$ in network's sum-rate corresponding to the lowest and highest traffic loads, respectively. 
\end{abstract}

\IEEEpeerreviewmaketitle

\section{Introduction}\label{sec:intro}
With the explosive bandwidth demand of wireless devices, milli-meter Wave (mmWave) is considered a promising solution to the spectrum scarcity problem. mmWave provides a large spectrum in the above-$6$ GHz band, i.e., Frequency Range 2 (FR-2). In contrast to sub-$6$ GHz band, FR-2 suffers from higher propagation losses that limit the coverage range of communication. Therefore, beamforming is used to combat mmWave losses by reshaping the beam pattern of the antenna in the direction of the user, hence achieving better power density in the direction of propagation. On the other hand, Non-Orthogonal Multiple Access (NOMA) is a promising multiple access technique for 5G and beyond 5G networks. The key idea in NOMA is to serve multiple users on the same time/frequency resources while superposing their messages in power domain, i.e. allocating different power levels to users' signals. This superposition process relies on the relative channel gains of the users such that users with better channel gains get less power levels; whereas users with bad channel gains get higher power levels. Successive Interference Cancellation (SIC) is applied at the users' side to remove inter-user interference. In particular, the user with the best channel decodes its message by successively decoding other users' messages and subtracting their effect from the received signal; whereas users with bad channel decode their respective signals directly \cite{6868214}. 

Despite the significant spectral efficiency and capacity improvements that the aforementioned techniques bring about, several challenges hinder that performance gain. In a downlink multi-beam scenario, the coverage of beams associated with different cells might intersect causing Inter-Beam Inter-Cell Interference (IB-ICI). Careful allocation of power to each beam, i.e. inter-beam power allocation, is essential for IB-ICI mitigation. Furthermore, the number of users covered by a single beam impacts the complexity and performance of SIC. As mentioned earlier, SIC performs successive decode and encode iterations to remove inter-user interference. Therefore, increasing number of users per beam leads to a large increase in complexity. Furthermore, the performance of SIC diminishes rapidly as the number of users increases \cite{6861434}. To prevent SIC performance degradation, hence improve sum rate, it is imperative to balance the load across cells through user-cell association. In parallel to advances arising from mmWave, beamforming and NOMA, there are significant efforts to make use of machine learning techniques to improve the performance of next-generation wireless networks \cite{8758918}.\\ 
In this paper, we address IB-ICI by using machine learning for joint user-cell association and inter-beam power allocation. In particular, we use a Q-learning algorithm that aims to enhance the sum rate of the network. Our results show that the proposed algorithm increases the achieved sum rate with at least $13\%$ for the least offered traffic load with a convergence of about $286$ ms. In addition, about $30\%$ increase in sum rate is achieved in the highest traffic load simulated.

This paper is organized as follows. Section \ref{sec:relWork} presents the related work. Section \ref{sec:sysModel} presents the system model and highlights the main trade-offs in maximizing the sum rate. In section \ref{sec:alg}, we provide a background on reinforcement learning and present the proposed algorithm which is based on Q-learning. Section \ref{sec:perfEval} presents the simulation setup, baseline algorithm and performance results. Finally, section \ref{sec:conclusion} concludes the paper. 

\section{Related Work}\label{sec:relWork}

In \cite{8454272}, the authors aim to maximize the sum rate of a mmWave NOMA system by solving user clustering and NOMA power allocation. The authors utilize the correlation features of the user channels to develop a k-means clustering algorithm. They also drive a closed form solution for the optimal NOMA power allocation within a cluster. The authors in \cite{8762180} aim to maximize the throughput of an ultra-dense mmWave network with multi-connectivity by solving the user-cell association problem. Using multi-label classification technique, the authors investigate three approaches for user-cell association: binary relevance, ranking by pairwise comparison, and random k-lebelsets. Besides these unsupervised learning techniques, reinforcement learning has been used in resource allocation in \cite{5GforumPaper} for capacity maximization and in \cite{8781859} for latency minimization.  

Beam selection and power allocation have been further studied in \cite{8782638} where the authors formulate a mixed integer non-linear programming problem for joint beam selection and power allocation in 5G mmWave small cell network. Due to the non-convexity of the problem, they decompose it to two sub-problems, i.e. beam selection and power allocation, and solve the former with an optimal algorithm based on cooperative games and the latter with Lagrange duality and non-cooperative games. Interference alignment techniques based on coordinated beamforming is used in \cite{7582424}, where two base stations jointly optimize their beamforming vectors to improve the downlink throughput of a multi-cell MIMO-NOMA network. In addition, the authors in \cite{Ishihara2017} address inter-cell interference mitigation in downlink multi-user MIMO for wireless LAN networks using transmit beamforming. The scheme relies on the estimation of both the power of the inter-cell interference and the channel state information to calculate the transmit beamforming weight matrix. 

The authors in \cite{7961156} formulate a multi-objective optimization problem for improving throughput and minimizing energy consumption in mmWave-based ultra-dense networks. The problem aims at solving the user association and power allocation with the constraints of load balancing, quality of service requirements, energy efficiency, energy harvesting, and cross-tier interference limits between macro base station and mmWave-based stations. 
The authors in \cite{8536429} consider a massive 5G femtocell network and design a cluster scheme to group femtocells and femto-users having the highest probabilities of line of sight connectivity. Inside each cluster, a joint user-association and resource allocation is performed. The authors use joint difference of two convex function to solve the clustering problem and subproblems of mixed interger non-linear programming to solve the user-cell association and power allocation. The authors in \cite{7802615} address the problem of user clustering, beamforming and NOMA power allocation in a downlink multi-user MIMO system. In particular, the authors propose a multi-clustering zero-forcing beamforming algorithm to mitigiate inter-cluster interference and maximize the sum spectral efficiency. Furthermore, they provide a dynamic power allocation solution to perform inter-cluster and intra-cluster power allocation. In \cite{7084118}, the authors consider a coordinated multi-cell downlink beamforming in massive MIMO systems with the aim of minimizing the aggregate transmit power of all base stations subject to Signal to Interference plus Noise Ratio (SINR) constraint. They propose two algorithms, a decentralized algorithm of coordinated beamforming using tools from random matrix theory, and a heuristic extension with base station transmit power constraint. Finally,  \cite{Attiah2019} presents a comprehensive survey on user-cell association and resource allocation in mmWave networks.

Unlike previous work, in this paper, we address inter-beam inter-cell and intra-beam interference using joint user-cell association and inter-beam power allocation. The main objective of our work is to improve network sum rate. To achieve this, we propose an online Q-learning algorithm with an action space of user-cell association and inter-beam power allocation. The proposed algorithm updates its decisions each scheduling interval, taking into account the interference among cells inferred from estimated SINR values at the UEs. 
\section{System Model}\label{sec:sysModel}
\textit{\textbf{Notations:}} In the remainder of this paper, bold face lower case characters denote column vectors, while non-bold characters denote scalar values. The operators $(.)^T$, $(.)^H$ and $|.|$ correspond to the transpose, the Hermitian transpose, and the absolute value, respectively. The operator $(A)^-$ under set $B$ represents the absolute complement of $A$, i.e. $(A)^- = B \setminus A$. 

\begin{figure}
    \centering
    \includegraphics[scale=0.35]{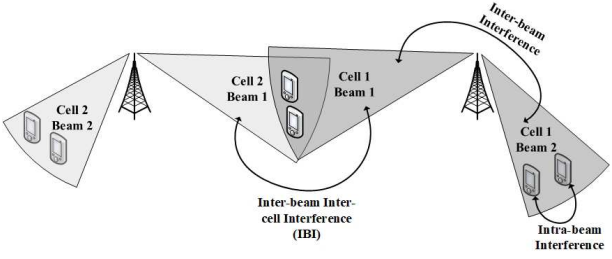}
    \caption{System model of mmWave network using beamforing.}
    \label{fig:sysModel}
\end{figure}

Consider a downlink mmWave-NOMA system with $J \in$ {\Fontauri\bfseries J} 5G-NodeBs (gNBs) equipped with $M \in$ {\Fontauri\bfseries M} transmit antennas and $U \in$ {\Fontauri\bfseries U} single-antenna users. Furthermore, users are partitioned into different clusters, $k \in$ {\Fontauri\bfseries K}, that are served using different beams such that {\Fontauri\bfseries U}$_k$ is the set of users covered by $k^{th}$ beam. Let {\Fontauri\bfseries K}$_j$ be the set of beams of $j^{th}$ gNB.  Henceforth, we use cluster and beam interchangeably. Indeed, different beams of different cells can have coverage intersection as shown in Fig. \ref{fig:sysModel}. Such intersection gives rise to Inter-Beam Inter-Cell Interference (IB-ICI). IB-ICI mitigation is essential in order to maximize network rate. 

In this paper, Poisson Cluster Process (PCP) is used to model users' deployment in the network, where the parent process follows a uniform distribution and the users of a cluster are uniformly deployed within a circular disk of radius $R_k$ around the cluster center. Every gNB performs a clustering algorithm to group users that can be covered by a single beam. Under every beam, downlink NOMA power allocation is used to multiplex users in the power domain, whereas users use SIC to demodulate their respective signals. We employ k-means clustering algorithm and the closed-form NOMA power allocation that is proposed in \cite{8454272}. In particular, k-means is used to cluster users according to the correlation of their wireless channel properties, i.e. users with correlated channels are more likely to be located close to each other. 

In mmWave channels, the gain of the Line-of-Sight (LoS) path is significantly larger than the gain of the Non-LoS (NLoS) path, i.e. with around $20$ dB \cite{8454272}, hence the mmWave channel model can be simplified to a single-path LoS model as follows: 
\begin{equation}
    \boldsymbol{h}_{k,u,j} = \boldsymbol{a}(\theta_{k,u,j}) \frac{\alpha_{k,u,j}}{\sqrt{L}(1 + d_{u,j}^{\eta})},
\end{equation}
where $L$ is the number of paths, $\boldsymbol{h}_{k,u,j} \in \mathbb{C}^{M \times 1}$ is the channel complex coefficient vector of $u^{th}$ user and $j^{th}$ gNB on $k^{th}$ beam, i.e. link $(u,k,j)$, $\alpha_{k,u,j} \in \mathbb{C}N(0, \sigma^2)$ is the complex gain, $d_{j,u}^{\eta}$ is the distance of link $(u,j)$ with pathloss exponent $\eta$. In addition, $\boldsymbol{a}(\theta_{k,u,j})$ is the steering vector, which can be represented as follows:
\begin{equation}
    \boldsymbol{a}(\theta_{k,u,j}) = [1, e^{-j 2 \pi \frac{D}{\lambda} \sin(\theta_{k,u,j})}, ..., e^{-j 2 \pi (M-1) \frac{D}{\lambda} \sin(\theta_{k,u,j})}]^T,
\end{equation}
where $D$ is the gNB's antenna spacing, $\lambda$ is the wavelength, $\theta_{k,u,j}$ is the Angle of Departure (AoD). 

\subsection{Problem Analysis}
In this work, we aim to improve the sum rate in mmWave network by performing user-cell association and inter-beam power allocation. In particular, sum rate can be calculated as follows: 
\begin{equation}
    R_{sum} = \omega \sum\limits_{j \in \text{\Fontauri\bfseries J}} \sum\limits_{k \in \text{\Fontauri\bfseries K}_j} \sum\limits_{u \in \text{\Fontauri\bfseries U}_k} \log_2(1 + \Gamma_{k,u,j}),
    \label{eq:sumRate}
\end{equation}
where $\omega$ is the bandwidth, and $\Gamma_{k,u,j}$ is the SINR of $(u,k,j)^{th}$ link, which can be expressed as:
\begin{equation}
    \Gamma_{k,u,j} = \frac{P_{k,j} \beta_{k,u,j} |\boldsymbol{h}_{k,u,j}^H \boldsymbol{w}_{k,j}|^2}{I_1 + I_2 + \sigma^2},
    \label{eq:sinr}
\end{equation}
\begin{equation}
    I_1 = P_{k,j} |\boldsymbol{h}_{k,u,j}^H \boldsymbol{w}_{k,j}|^2 \sum\limits_{\substack{i \neq u \\ O(i) > O(u)}} \beta_{k,i,j},
    \label{eq:inter1}
\end{equation}
\begin{equation}
    I_2 = \sum\limits_{l \in \text{\Fontauri\bfseries K}_{(j)^-}} P_l |\boldsymbol{h}_{l,u,(j)^-}^H \boldsymbol{w}_{l,(j)^-}|^2,
    \label{eq:inter2}
\end{equation}
where $P_{k,j}$ denotes the power allocated to $k^{th}$ beam of $j^{th}$ gNB, and $P_l$ is the power allocated to $l^{th}$ interfering beam. $\beta_{k,u,j}$ and $\beta_{k,i,j}$ is the power allocation factor of $(k,u,j)^{th}$ and $(k,i,j)^{th}$ links respectively. $\boldsymbol{w}_{k,j}$ is the beamforming vector, and $\sigma^2$ represents receiver's noise variance. The setup shown in Fig. \ref{fig:sysModel} presents three types of interference: Intra-beam interference, IB-ICI, and inter-beam interference. In this paper, different beams are allocated different spectrum bands, hence inter-beam interference becomes void. With NOMA power allocation, users sharing the same time/frequency resources, are multiplexed in the power domain. This incurs intra-beam interference as expressed in Eq. \ref{eq:inter1}. Finally, IB-ICI is expressed in Eq. \ref{eq:inter2}. $O(u)$ denotes the decoding order of $u^{th}$ user whereas $\text{\Fontauri\bfseries K}_{j^-}$ denotes the set of beams that belong to the absolute complement of $j$ under set {\Fontauri\bfseries J}, i.e. $((j)^- = \text{\Fontauri\bfseries J} \setminus j)$. Finally, $\boldsymbol{h}_{l,u,(j)^-}$ represents the channel vector between the $l^{th}$ interfering beam from other cells in the set $(j)^-$ and $u^{th}$ user, and $\boldsymbol{w}_{l,(j)^-}$ is the beamforming vector of $l^{th}$ interfering beam. 



\section{Proposed Machine Learning Approach}\label{sec:alg}

\subsection{Background on Q-learning}
Q-learning algorithm is a sub-class of reinforcement learning algorithms. Reinforcement learning refers to any agent-oriented learning, in which an agent interacts with an environment towards achieving a certain goal. In particular, the agent aims to learn the dynamics of the environment through trial and error. In response to the agent's actions, the agent receives a quantitative feedback that represents either a reward or a cost and the environment's state changes. Indeed, such setup can be represented as a Markov Decision Process (MDP) with a tuple of (agents, states, actions, and reward function). 
The ultimate goal of the agent is to maximize the total expected future discounted rewards. To achieve that, the agent aims to find a policy that quantifies the optimal action decision in each state. This can be done using an action-value function as follows:
\begin{equation}
    q_{\pi}(s, a) = \mathbb{E}_{\pi}[R_{t+1} + \gamma R_{t+2} + \gamma^2 R_{t+3}... | S_t = s, A_t = a],
\end{equation}
where $q_{\pi}(s, a)$ is the quality value, i.e. Q-value, of policy $\pi$ when starting at $s^{th}$ state and taking $a^{th}$ action. In particular, the optimal Q-values can be computed using a brute-force method. However, in order to facilitate online learning, an iterative algorithm, Q-learning, is used to approximate Q-values each iteration. Q-learning is a temporal difference method that uses the following update to approximate an agent's policy:
\begin{equation}
    q(s, a) \gets q(s, a) + \alpha [R + \gamma \max\limits_a q(s', a) - q(s, a)],
    \label{eq:qvalue}
\end{equation}
where $R$ is the reward value, and $\max\limits_a q(s', a)$ computes an approximate of the Q-value at the next state $s'$. 
In the next section, we present the proposed Q-learning algorithm to perform joint user-cell association and inter-beam power allocation for maximizing network's sum rate. 
\subsection{Proposed Q-learning Algorithm}

We define an online multi-agent Q-learning algorithm as follows:
\begin{itemize}
    \item \textbf{Agents:} gNBs.
    \item \textbf{Actions:} Each gNB decides on its user associations and inter-beam power allocation. The user-cell association is performed only for users that lie in the intersection region of two or more cells. Let {\Fontauri\bfseries U}$_j^{int}$ be the set of users of $j^{th}$ gNB that lie in its intersection region with other cells. The vector of actions is defined as $\boldsymbol{a}_j = [\boldsymbol{\delta_{j}}, \boldsymbol{P_j}; j \in \text{\Fontauri\bfseries J}]$, where $\boldsymbol{a}_j \in A_j$. The vector $\boldsymbol{\delta_{j}} = [\delta_{j,i}, i \in \text{\Fontauri\bfseries U}_j^{int}]$ represents a binary vector of user-cell association where each element indicates whether the gNB decides to associate the $i^{th}$ user, $\delta_{j,i} = 1$, or not, $\delta_{j,i} = 0$. Furthermore, $\boldsymbol{P_j} = [P_{j,k}, k \in \text{\Fontauri\bfseries K}_j]$ represents a vector that defines the power allocated to each beam of $j^{th}$ gNB. As such the size of the action-space becomes $2^{|\text{\Fontauri\bfseries U}_j^{int}|} \times N_p^{|\text{\Fontauri\bfseries K}_j|}$, where $N_p$ represents the set of power values available for each beam.
    \item \textbf{States:} We define the states in terms of the average SINR which reflects the level of interference in the wireless environment:
    \begin{equation}
        S_{j} = 
        \begin{cases}
            S_0 \tabmore \overline{\Gamma}_j \geq \Gamma_{th}, \\
            S_1 \tabmore otherwise,
        \end{cases}
        \label{eq:states}
    \end{equation}
    where $j^{th}$ gNB, i.e. agent, transits to state $S_0$ as long as its average SINR, $\overline{\Gamma}_j$ is greater than a threshold value, $\Gamma_{th}$, and transits to $S_1$ otherwise. The average SINR of $j^{th}$ gNB is defined as follows:
    \begin{equation}
        \overline{\Gamma}_j = \frac{1}{(K \times U)} \sum\limits_{k \in \text{\Fontauri\bfseries K}_j} \sum\limits_{u \in \text{\Fontauri\bfseries U}_k} \Gamma_{k,u,j}
    \end{equation}
    \item \textbf{Reward:} We formulate the reward function based on SINR as follows:
    \begin{equation}
        R_j = 
        \begin{cases}
        1 \tabmore \overline{\Gamma}_j \geq \Gamma_{th}, \\
        -1 \tabmore otherwise,
        \end{cases}
        \label{eq:reward}
    \end{equation}
\end{itemize}
\begin{algorithm}
    \begin{algorithmic}[1]
    	\STATE \underline{\textbf{Initialization:}} Q-table $\gets$ 0, $\alpha$, $\gamma$, and $\epsilon$.
        \FOR{scheduling assignment period $t$ = 1 to $T$}
          \STATE \underline{\textbf{Step 1:}} Receive SINR estimations from attached users.  
          \STATE \underline{\textbf{Step 2:}} Perform Q-learning algorithm for joint user-cell association and inter-beam power allocation:
          \bindent
            \STATE Compute average SINR of users in the intersection region.
            \STATE Update reward as in Eq. (\ref{eq:reward}). 
            \STATE Update Q-value (and Q-table) as in Eq. (\ref{eq:qvalue}).
            \STATE Switch to state $s'$ as in Eq. (\ref{eq:states}).
            \IF{rand $\leq \epsilon$}
                \bindent
                    \STATE $\boldsymbol{a}_j \gets \textit{draw uniformly from } A_j$
                \eindent
              \ELSE
                \bindent
                    \STATE $\boldsymbol{a}_j = \max\limits_{a \in A_j} q(s', a)$
                \eindent
              \ENDIF
          \eindent
          \STATE \underline{\textbf{Step 3:}} Downlink transmission of user-cell association decisions to each UE. 
          \STATE \underline{\textbf{Step 4:}} Wait UEs to perform final user-cell association decisions as in Algorithm \ref{alg:propAlg2}. 
          \STATE \underline{\textbf{Step 5:}} Receive final user-cell association decisions from UEs. 
          \STATE \underline{\textbf{Step 6:}} Perform k-means clustering and NOMA intra-beam power allocation. 
          \STATE \underline{\textbf{Step 7:}} Perform downlink transmission, while each user performs downlink reception using SIC.
        \ENDFOR
    \end{algorithmic}
    \caption{Proposed Q-learning algorithm for joint user-cell association and inter-beam power allocation (gNB)}
    \label{alg:propAlg}
\end{algorithm}
\begin{algorithm}[t]
    \begin{algorithmic}[1]
    \FOR{scheduling assignment period $t$ = 1 to $T$}
          \STATE \underline{\textbf{Step 1:}} Receive association decisions from gNBs. 
          \STATE \underline{\textbf{Step 2:}} Update priority list, i.e. maintain gNBs that decided to associate and remove gNBs that decided not to associate with the UE.
          \STATE \underline{\textbf{Step 3:}} Select the gNB with the highest priority on the list to associate with and send the final decision to the selected gNB. 
    \ENDFOR
    \end{algorithmic}
    \caption{User-cell association (UE)}
    \label{alg:propAlg2}
\end{algorithm}
\textit{Algorithm \ref{alg:propAlg}} presents the steps performed by each gNB, whereas \textit{Algorithm \ref{alg:propAlg2}} presents the steps performed by each user. Furthermore, user-cell association process involves Q-learning part at gNB's side and priority list at user's side. In particular, the user maintains a priority list of the gNBs to associate with, which is computed according to SINR estimation in the last transmission interval. Afterwards, each gNB performs the Q-learning algorithm which results in an association decision for each user in the intersection region, where each user is informed about that decision. Finally, each user follows \textit{Algorithm \ref{alg:propAlg2}} to combine the decisions from gNBs with its priority list and informs the selected gNB. 

\section{Performance Evaluation}\label{sec:perfEval}

\subsection{Simulation Setup}
We use 5G Matlab Toolbox to construct a discrete event simulator. The simulator works on TTI level with 5G downlink transmission and reception. Table \ref{tab:simSettings} presents the simulation settings. The network is composed of two gNBs with inter-gNBs distance of $150$ m. Users are stationary and their positions follow a PCP with $\lambda = 7$. We consider $2$ clusters, and cluster radius of $30$ m. The performance of the proposed algorithm is tested under several traffic loads. The number of users in the intersection region is $2$, number of power levels used is $5$, number of clusters is $2$, and number of states is $2$. Hence, the size of the action-space becomes $2^2 \times 5^2 = 100$ and the size of the Q-table is $2^2 \times 5^2 \times 2 = 200$. In addition, we employ k-means clustering and closed-form NOMA power allocation proposed in \cite{8454272} as a base for our implementation. 

The proposed algorithm is compared to a baseline algorithm that heuristically performs user-cell association and inter-beam power allocation. In particular, the baseline algorithm performs user-cell association by constructing a priority list of gNBs ordered according to SINR. Afterwards, users associate with the gNB with the highest priority on the list. In addition, power allocation is performed by equally dividing the total power of cell among its beams. 

\subsection{Performance Results}
In this section, we present performance results of the proposed Q-learning algorithm compared with the baseline algorithm, UPA in terms of sum rate, latency, and Packet Drop Rate (PDR). 

\begin{figure}[t!]
    \centering
    \includegraphics[scale=0.3]{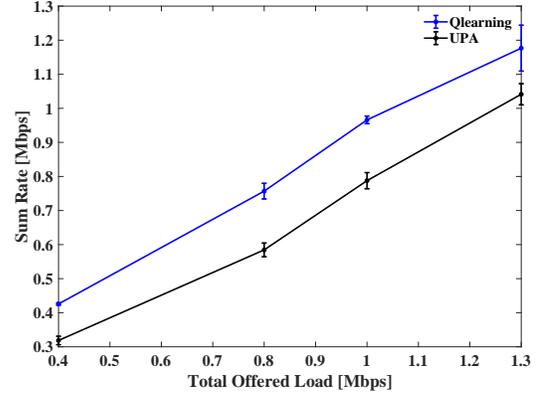}
    \caption{Sum rate [Mbps] versus total offered load [Mbps]. Number of users is 9.}
    \label{fig:sumRate}
\end{figure}

\begin{figure}[t!]
    \centering
    \includegraphics[scale=0.3]{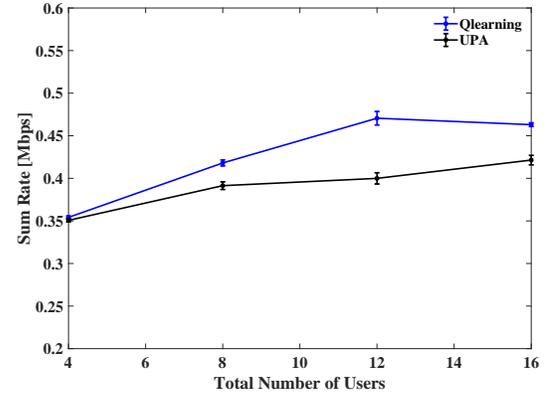}
    \caption{Sum rate [Mbps] versus total number of users with $0.5$ Mbps total offered load.}
    \label{fig:sumrateUsers}
\end{figure}

\begin{figure}[t!]
    \centering
    \includegraphics[scale=0.3]{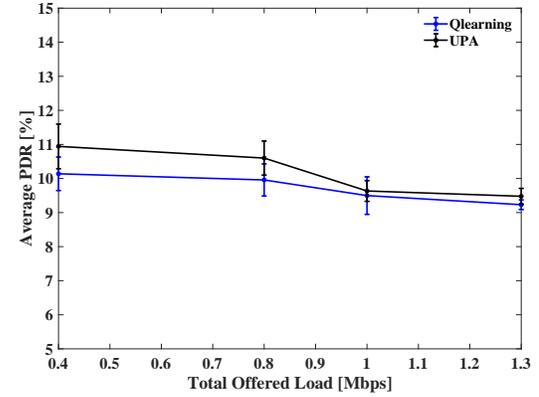}
    \caption{Average packet drop rate [\%] versus total offered load [Mbps]. Number of users is 9.}
    \label{fig:avgPDR}
\end{figure}

Fig. \ref{fig:sumRate} presents the network sum rate versus the total offered load. The figure shows that the proposed scheme outperforms UPA in all cases with a rate increase of $13\%$ and $33\%$ at the lowest and highest offered loads, respectively. In addition, Fig. \ref{fig:sumrateUsers} presents the network sum rate versus the total number of users in the network. The figure shows that Q-learning is able to maintain a sum rate close to the total offered load (which is set to 0.5 Mbps for the presented case) when increasing the number of users per network, whereas UPA is achieving lower sum rate. The packet drop rate (PDR) is presented in Fig. \ref{fig:avgPDR}, where both algorithms are achieving very comparable PDR (around $10-11\%$).
\begin{table}[htp]
    \centering
    \caption{5G mmWave Network Simulation Settings}
    \begin{tabular}{|l|l|}
         \hline
         \textbf{\underline{5G PHY configuration}} & \\
         Bandwidth & $20$ MHz \\
         Carrier frequency & $30$ GHz \cite{8782638}\\
         Subcarrier spacing & $15$ KHz \\
         Subcarriers per resource block & $12$ \\
         TTI size & $2$ OFDM symbols ($0.1429$ msec) \\
         Max transmission power & $28$ dBm \\
         \hline
         \textbf{\underline{HARQ}} & \\
         Type & Asynchronous HARQ \\
         Round trip delay & $4$ TTIs \\
         Number of processes & $6$ \\
         Max. number of re-transmission & $1$ \\
         \hline
         \textbf{\underline{Distribution of users}} & \\
         Mobility & Stationary \\ 
         Distribution & Poisson Cluster Process (PCP) \\
         PCP Average number of users & $7$ \\
         Number of clusters & $2$ \\
         Radius of cluster & $30$ m \\
         Number of users & $4-16$ \\
         Number of gNBs & $2$ \\
         Inter-gNBs distance & $150$ m \cite{8782638} \\
         \hline
         \textbf{\underline{Traffic}} & \\
         Distribution & Poisson \\
         Packet size & $32$ Bytes \\
         \hline
         \textbf{\underline{Q-learning}} & \\
         Learning rate $(\alpha)$ & $0.5$ \\
         Discount factor $(\gamma)$ & $0.9$ \\
         Exploration probability $(\epsilon)$ & $0.1$ \\
         Inter-beam power levels & $[0:2:8]$ dBm \\
         Threshold SINR ($\gamma_{th}$) & 20 dB \\
         \hline
         \textbf{\underline{Simulation parameters}} & \\
         Simulation time & $4000$ TTI \\
         Number of runs & $40$ \\
         Confidence interval & $95\%$ \\
         \hline
    \end{tabular}
    \label{tab:simSettings}
\end{table}

\begin{figure}[t!]
    \centering
    \includegraphics[scale=0.3]{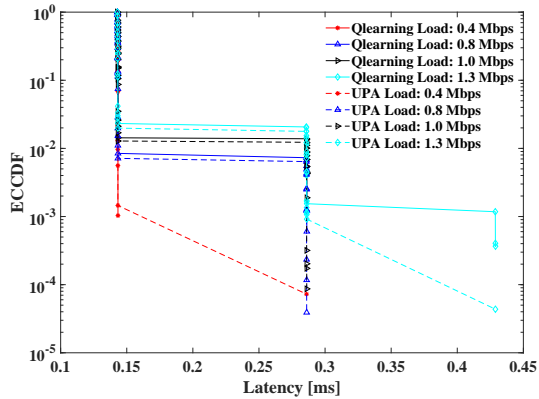}
    \caption{Average Latency [ms] versus Total Offered Load [Mbps].}
    \label{fig:avgLatency}
\end{figure}

Furthermore, Fig. \ref{fig:avgLatency} shows the empirical Complementary Cumulative Distribution Function (eCCDF) of the average achieved latency. Latency is defined as the delay of the packet since its creation at the gNB until its delivery at the user side. This includes queuing, transmission, and propagation delays. The processing at both ends, i.e. gNB and user, includes RLC, MAC and PHY layers. The figure presents that both algorithms achieve similar latency values at different offered loads. The figure also shows three main latency points: $0.1429$ ms, $0.2857$ ms, and $0.4286$ ms, which correspond to 1, 2, and 3 TTIs respectively, where 1 TTI represents 2 OFDM symbols. In particular, queuing and re-transmission delays contribute to the total achieved delay \cite{medhatelsayed1}. By improving interference, i.e. SINR, re-transmission delay, hence total latency improves. 

\begin{figure}[t!]
    \centering
    \includegraphics[scale=0.3]{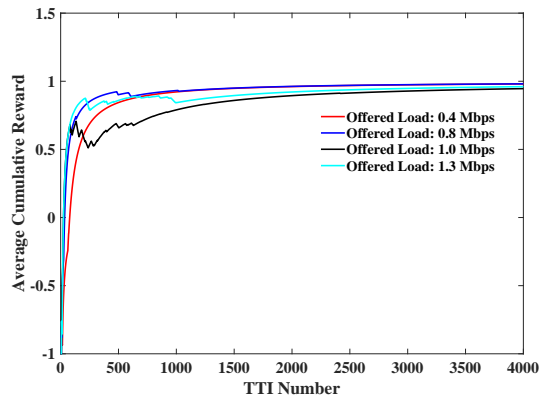}
    \caption{Cumulative Average of Q-learning's Reward versus Iteration Number with different Total Offered Load.}
    \label{fig:convergenceAll}
\end{figure}

Finally, Fig. \ref{fig:convergenceAll} shows the average cumulative reward versus the iteration number. The $\epsilon$-greedy action selection methodology, presented on lines 9-13 in Algorithm \ref{alg:propAlg}, is applied for $2000$ TTIs, whereas greedy policy is followed afterwards. The proposed algorithm converges at around the $2500^{th}$ TTI with a slight decrease of the reward at $500-600^{th}$ TTI due to the exploration policy. 

\section{Conclusion}\label{sec:conclusion}
In this paper, we presented a machine learning algorithm to address the joint problem of user-cell association and inter-beam power allocation in 5G mmWave network. The proposed algorithm aims at improving the sum rate by mitigating the intra-beam interference and inter-beam inter-cell interference. On one hand, the algorithm performs inter-beam power allocation such that it balances the interference posed by beams of adjacent cells. On the other hand, the algorithm performs user-cell association which balances users' attachments across cells, hence improving the performance of successive interference cancellation. The proposed algorithm is designed using an online Q-learning algorithm and compared with a baseline algorithm that uses uniform power allocation. Simulation results reveal the ability of the proposed algorithm to improve the sum rate. 
\section*{Acknowledgment}
This research is supported by the Natural Sciences and Engineering Research Council of Canada (NSERC) under Canada Research Chairs Program and CREATE program. 

\bibliographystyle{ieeetr}
\bibliography{reference}

\end{document}